# Acoustic Vortex Filter Based on Tunable Metasurfaces

Liulin Li,[1] Bingyi Liu,[1, a)] Zhixiang Li,[1] Kai Guo,[1] and Zhongyi Guo[1, a)]

**AFFILIATIONS**

[1]School of Computer Science and Information Engineering, Hefei University of Technology, Hefei, 230009, China.

[a)]**Author to whom correspondence should be addressed:** *bingyiliu@hfut.edu.cn, guozhongyi@hfut.edu.cn*

## ABSTRACT

In this paper, we present an acoustic vortex filter (AVF) based on tunable metasurfaces, which can selectively filter the incident multiplexed vortices that carry different orbital angular momentum (OAM). Our metasurface-based AVF is composed of an upper acoustic metasurface (UAM) and a lower acoustic metasurface (LAM), of which the intrinsic topological charge (ITC) can be tuned by mechanically rotating the UAM along its central axis. Due to the critical order of the propagating vortex modes in waveguide, controlling the ITC of the AVF allows for the selective filtering of incoming multiplexed acoustic vortex beams based on the sound vortex diffraction in phase-gradient metasurface, which endows the vortex filter the capability that let the incident vortex of specific OAM pass through it. In the following demonstration, both in theory and experiment, we design the AVF and effectively filter the acoustic vortices with two opposite topological charges (TCs) by simply altering the orientation angle of the UAM. Based on this, we further demonstrate its application in asymmetric acoustic wave transmission. Our work offers an approach to selectively filter the incident acoustic vortex, which improves the capability to control the acoustic OAM via metasurfaces.

As acoustic wave is the only candidate for long-range underwater communication, the capacity of traditional acoustic communication channel is inherently limited by low operating frequency and low speed, so how to further improve the bandwidth of acoustic communication has become an urgent problem to be solved.[1-5] In recent years, vortex beams carrying orbital angular momentum (OAM) provide another degree of freedom (DoF) besides the common DoFs like amplitude, phase, or frequency. In theory, there are infinite orthogonal OAM modes, which can effectively alleviate the capacity crisis in the field of free-space communication.[6-10] Therefore, OAM-based acoustic communication technique has attracted tremendous research interests in recent years.

The vortex beam carrying specific OAM is typically characterized by a spiral wavefront of $\exp(il\theta)$, where its phase factor is linearly proportional to the azimuthal angle $\theta$. And the integer $l$, which is known as the topological charge (TC) or OAM mode number, is the number of the times wavefront rotates over a wavelength distance, and it characterizes the magnitude and direction of beam's OAM.[11-13] Moreover, a phase singularity appears at the beam's center, which gives its field intensity a donut shape. Besides, the vortex beams carrying different TCs are orthogonal to each other, allowing the efficient multiplexing and demultiplexing of multiple data channels.[14-16] To date, thanks to the intriguing characteristics of vortex beams, they are widely applied to the field of acoustic communication,[17-21] information processing[22-23], acoustic torque[24,25] and particle trapping[26-27], etc.

At present, the main solutions for acoustic vortex generation can be divided into two categories, i.e., active, and passive technique. Phased array consisting of individually addressable transducers of designated initial phase and amplitude is a classical active method to generate the OAM of desired TC. But advanced digital control of numerous transducers requires complex electronic circuits of high cost. In recent years, acoustic metasurfaces have been widely employed in acoustic wavefront manipulation, providing a promising approach to acoustic vortex beam generation.[28-31] However, the functionality of acoustic meta-atoms is fixed once the geometry of the structure is determined. Therefore, how to generate multiple acoustic vortex beams with simple and reconfigurable metasurfaces has become a research hotspot.[32-34]

It should be noted that, besides the generation of acoustic OAM, the subsequent manipulations of OAM are also essential, especially the emerging concept of twist acoustics, where complex acoustic vortex manipulation is realized with multiple acoustic structures, such as acoustic geometric phase,[35-38] acoustic vortex absorption,[39] vortex assisted acoustic frequency conversion,[40] etc. In addition, OAM detection or filtering is also important in the OAM-based communication system as an example, the transformation and manipulation of different channels plays an important role in improving its flexibility and practicality.[41] According to the application scenarios, users may have customized requirements for the channels where different OAM beams reside, especially selecting a specific channel in multiplexed OAM-based communication links. Therefore, it is important to develop the OAM-filter technology that filters out useless channels and selectively extract specific channels on demand.

In this paper, we propose a dual-layer metasurface-based tunable acoustic vortex filter (AVF) for the selective extraction and filtering of the acoustic multiplexed vortices. Our AVF could tune its intrinsic topological charge (ITC) $l^\xi$ by simply rotating its constituent elements. When OAM beams of different TCs are normally incident on the AVF, the selection of ITC $l^\xi$ will significantly influence its transmission property according to the sound vortex diffraction law, which is possible to prevent the vortex of selected TC passing through the AVF and thereby achieving the OAM filtering functionality. In this work, we theoretically and experimentally demonstrate the selective filtering of incident multiplexed vortex beams of two opposite TCs by simply adjusting the ITC of the AVF via mechanical rotation. In addition, we further verify the asymmetric vortex beam transmission realized with our AVF. In this scenario, our proposed AVF can block most of the energy of the incident vortex with the transmission loss contrast greater than 3 dB from two sides over a 160 Hz frequency bandwidth.

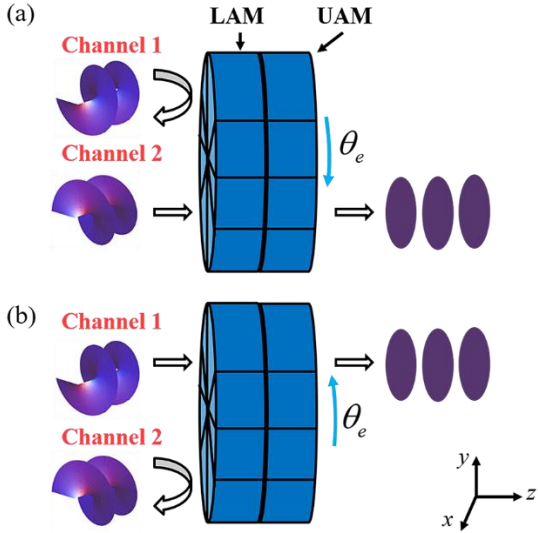

FIG. 1. Conceptual schematic of the AVF for selective filtering of acoustic vortex carrying specific OAM. When UAM of AVF is clockwise or counterclockwise rotated by angle of $\theta_e$, two filtering functionalities are respectively realized, i.e. (a) vortex of TC +1 transmits and vortex of TC –1 reflects at the orientation angle $\theta_e$; (b) vortex of TC +1 reflects and vortex of TC –1 transmits at the orientation angle $-\theta_e$.

Fig. 1 schematically shows the concept of AVF that selectively filters the incident multiplexed vortices by simply rotating its constituent component towards two opposite directions. In general, a reconfigurable AVF consists of two parallel layers: an upper acoustic metasurface (UAM) and a lower acoustic metasurface (LAM). In our design, each layer is composed by $m$ sectors of equal sector angle. By rotating the UAM counterclockwise or clockwise by angle of $\theta_e$, the AVF becomes a phase-gradient metasurface (PGM) and can generate vortex beam under normal acoustic plane wave illumination. It is well known that in a cylindrical waveguide of radius $R$, there exists finite vortex modes for a given operating frequency. Here, the maximum TC, i.e., the critical order $l_M$ specifies the range of TCs or orders of the propagating vortex beams from $-l_M$ to $+l_M$, where "+" and "–" correspondingly refer to counterclockwise and clockwise helicity of the propagating vortex (facing the propagating direction, i.e., +z direction given in Fig. 1). When an acoustic vortex beam of TC $l_{in}$ is normally incident on the AVF of ITC $l^\xi$, the transmitted vortex beam of TC $l_{out}^t$ can be obtained due to the sound vortex diffractions. The generalized conservation principle between the input and output TC of the vortex beams is known as the sound vortex diffraction law:[42]

$$l_{out}^{t(r)} = l_{in} + nl^\xi . \quad (1)$$

Here, the superscript $t$ and $r$ refers to the transmission and reflection mode, and $n$ is the diffraction order. Eq. (1) can be regarded as the grating equation or the generalized Snell's law in cylindrical coordinate, and $n$ is generally selected as 1.[43-47] Therefore, a critical order $l_C$ exits when $|l^\xi| \leq 2l_M$, and $l_C = \text{sgn}(l^\xi)(l_M - |l^\xi|)$, here "sgn" refers to sign function. For the convenience of the following discussion, we assume $l^\xi$ to be a positive integer. If the TC of incident vortex is within the critical order, i.e., $l_{in} \in [-l_M, l_C]$, then $|l_{out}| \leq l_M$ holds true, and the transmitted wave will be a propagating mode. Otherwise, when the TC of incident vortex is beyond critical order, the transmitted vortex mode is not allowed to propagate in the waveguide, and its longitudinal wave vector $k_z$ becomes a pure imaginary value becomes an evanescent wave. However, such process can be broken by high-order diffraction and gives the anomalous transmission or reflection of vortex, which is determined by the parity of term $L = m + n$. The physics meaning of $L$ is the round-trip times of the energy trapped in the PGM, which explicitly gives: [42,48]

$$\begin{aligned} l_{out}^t &= l_{in} + nl^\xi \ (L \text{ is odd}) \\ -l_{out}^r &= l_{in} + nl^\xi \ (L \text{ is even}) \end{aligned} \quad (2)$$

Therefore, it is possible to construct a vortex filter that selectively passes one vortex but totally blocks another vortex based on the PGM of proper ITC and parity. In other words, we could set our AVF to selectively transmit the vortex whose TC locates in $[-l_M, l_C]$ while the vortex whose TC locates in $[l_C, l_M]$ is reflected. When flipping the ITC of AVF made of reconfigurable metasurfaces, the domain of allowed input TC would also flip. In this case, we could selectively filter the incident vortex by setting the ITC of AVF via simple rotation of UAM.

To demonstrate such AVF driven by the sound vortex diffraction law, we take the case of $l_M = 1$, $l^\xi = \pm 1$ for illustration. For the convenience of analysis, we theoretically study the AVF made of ideal acoustic materials, i.e., subwavelength acoustic waveguides filled with the homogeneous acoustic medium of designed refractive index and impedance that is matched with the background medium (air in our work). The transmitted phase modulation of each sector is achieved by introducing corresponding refractive index modulation. For example, for the cell unit of height $h = 0.5\lambda$ with external phase modulation of $\Delta\varphi$, it satisfies $\Delta\varphi = k_0 h \Delta n$, here $k_0 = 2\pi/\lambda$ is the wave number of the acoustic wave in the air, $\lambda$ is the operating wavelength. Therefore, the external refractive index modulation of each sector can be calculated with $\Delta n = \Delta\varphi / k_0 h$. Then the required phase shift of each sector of the two layers of AVF can be determined according to our previously published work,[32] see supplementary material for more details.

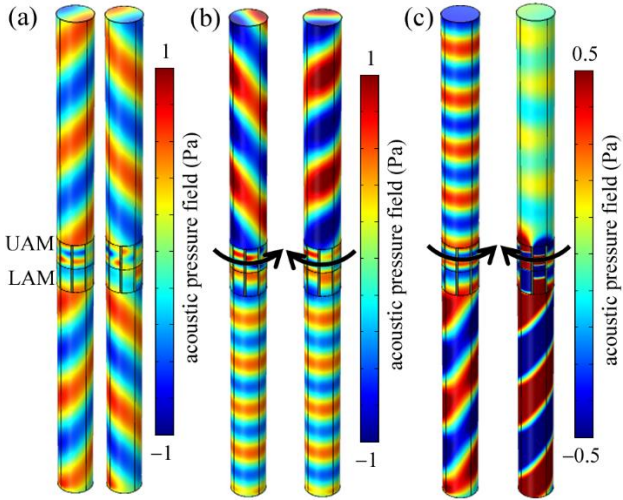

FIG. 2 Acoustic pressure field of the ideal AVF made of 8 sectors. Vortex incidence while no rotation of the UAM, the TC of incident vortex is (a) 1 and −1. Plane wave incidence while the UAM is rotated by (b) 45° and −45°. Vortex incidence while the rotation angle of UAM is (c) 45° and −45°.

Firstly, we investigate the diffraction property of the AVF made of 8 sectors. Fig. 2(a) shows the acoustic pressure field when vortex of TC ±1 is illuminated on the AVF without any external rotation, where we could observe direct transmission of the vortex, in this case, the ITC of the AVF is 0. Fig. 2(b) shows the acoustic pressure field when the AVF is illuminated by a plane wave, and here the UAM of AVF is rotated by angle of 45° or −45°. Based on the TC of the transmitted vortex field, we could accordingly determine the ITC of the AVF. In this case, when UAM is rotated by 45°, the ITC of the AVF becomes −1; when UAM is rotated by −45°, the ITC of the AVF becomes 1. Fig. 2(c) shows the acoustic pressure field when vortex of TC 1 is illuminated on the AVF. When the UAM is rotated by angle of 45°, the ITC of the AVF becomes −1. In this case, the diffraction order $n = 1$ is allowed and the transmission field is a plane wave whose TC is $l_{out}^t = 0$. When UAM is rotated by angle of −45°, its ITC becomes 1, and the allowed maximum diffraction order is $n = -2$. In this case, $L$ is 6 and its parity is even, according to Eq. (2), no transmission is expected, and a reflected vortex of TC $l_{out}^r = 1$ is observed. Therefore, Fig. 2(c) exactly shows a tunable vortex filter we intend to realize, i.e., for a given vortex, its transmission property can be switched by altering the rotation angle of UAM.

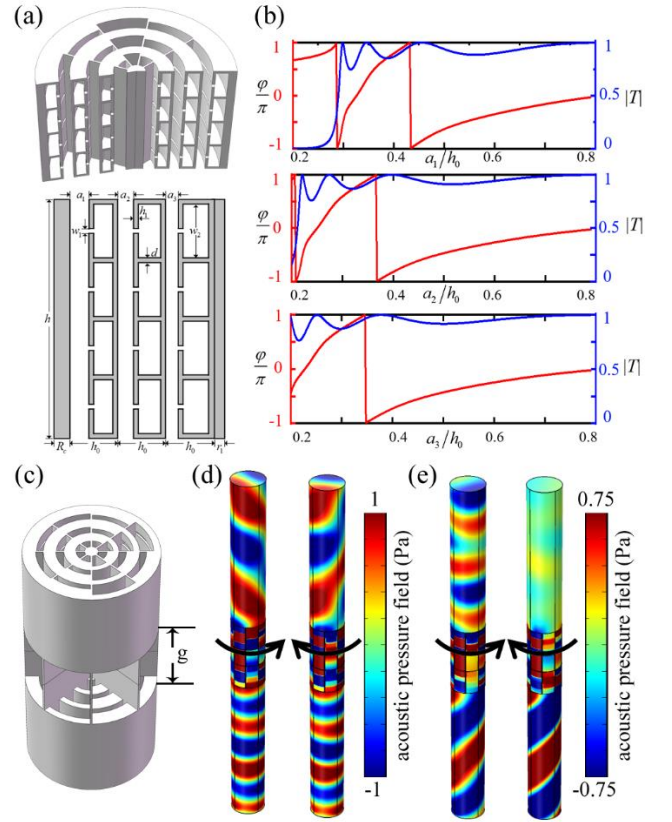

FIG. 3 (a) Schematic of Helmholtz resonator-based sector unit. (b) the phase (red) and amplitude (blue) of the transmission coefficient as functions of $a_i/h_0$ simulated for the $i$-th layer. (c) Schematic of AVF made of 8 sectors with external propagating distance $g$=55 mm among the UAM and LAM. (d) Acoustic pressure field when the AVF is illuminated by a plane wave, and the rotation angle of UAM is ±45°. (e) Acoustic pressure field when the AVF is illuminated by vortex of TC 1, and the rotation angle of UAM is ±45°.

To verify the tunable AVF functionality with realistic structures, we take advantage of the classical hybrid type acoustic structure as the building block, which is made of four cascade Helmholtz resonators and a straight pipe (HR-SP). The radius of waveguide is set as 5 cm and the operating wavelength $\lambda$ is 15 cm. The sections that make up the metasurface are composed of three layers HR-SPs, see Fig. 3(a). Parameters $h_1$, $d$, $w_1$, and $w_2$ are the neck height, thickness of the rigid wall, neck width, and width of the HR cavity, respectively. The period of each structure layer along the radial direction is $h_o$. To facilitate the assembly of each sector and fit the outer waveguide, a central cylinder of radius $R_C$ and an outer layer of thickness $r_l$ are supplemented in the design. To be more specific, the values of above parameters are given as follows: $R_C$ = 0.5 cm, $h_o$ = 1.5 cm, $h$ = 7.5 cm, $h_1$ = 0.15 cm, $w_1$ = 0.18 cm, $w_2$ = 1.6 cm, $d$ = 0.15 cm and $r_l$ = 0.3 cm. In addition, $a_1$, $a_2$ and $a_3$ are the width of the open pipe of each layer. By adjusting the above three parameters, we can achieve the flexible phase regulation and high transmission, and the relationship between the parameter $a_i$ $(i=1,2,3)$ and the transmission coefficient is illustrated in Fig. 3(b). Generally, the distance between UAM and LAM should be zero. However, in our study, we find that the interaction among the units of UAM and LAM would distort the overall phase delay and the transmittance of top and bottom units. The error induced by the interlayer coupling would make the diffraction property of AVF not consistent with the sound vortex diffraction law, such contradiction is attributed to the fact that some sectors fail to work properly, see supplementary material for more details. Fortunately, an effective solution is to introduce external propagating distance among the cascaded two units. Next, we perform the numerical calculations with the finite element methods to theoretically verify the selective filtering of the vortex beams by our AVF. Fig. 3(c) shows the schematic of AVF made of 8 sectors while introducing an external propagation distance $g$ = 55 mm between the UAM and LAM. Fig. 3(d) shows the ITC of the AVF is $ml$ when it is rotated by angle of ±45°. In our theoretical study, the incident vortex $l_{in}$ = 1 is obtained with four phased transducers of 0, $\pi/2$, $\pi$, $3\pi/2$ initial phase delay.[37] Fig. 3(e) shows the case when the vortex of TC 1 is illuminated on the AVF of ITC $\mp 1$, which is realized by rotating the UAM by angle of ±45°. Here we could obtain a plane wave transmission and a reflected vortex of TC 1. Therefore, the AVF made of realistic structures could support good vortex filter performance.

Interestingly, we design the AVF made of 9 sectors and also obtain the vortex filter functionality. Here, no external gap is required which makes the AVF be more compact. The reason that why our AVF made of 9 sectors could possess its vortex filter functionality is attributed to the fact that two adjacent units possess very close phase delay due to the interlayer coupling, which makes the AVF an equivalent PGM made of 8 sectors, and its vortex filtering function thereby exists according to the sound vortex diffraction law. If we introduce a gap g = 5 mm, the AVF exactly operates like a PGM made of 9 sectors again, and no vortex filter function can be achieved, see supplementary material for the relevant discussions.

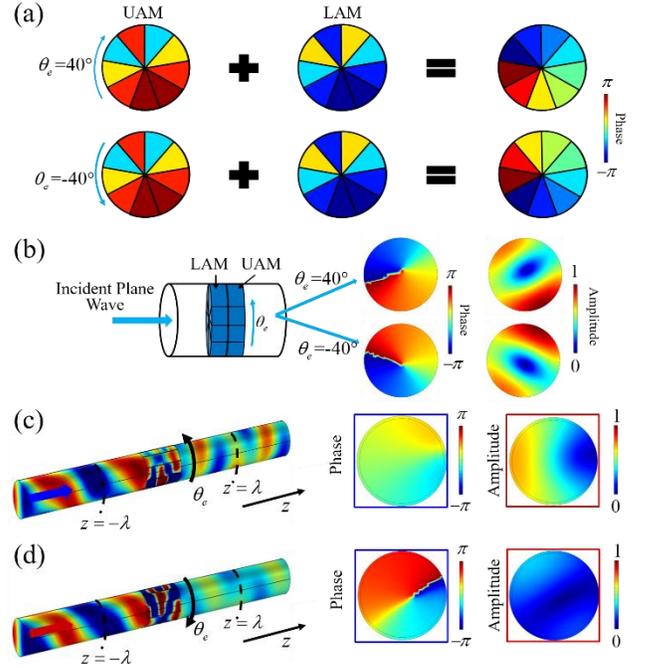

FIG. 4 Numerical demonstration of the selective vortex filtering by AVF made of 9 sectors and no gap is introduced. (a) Phase profile of the AVF formed by UAM and LAM when they are rotated to two opposite directions. (b) Acoustic pressure field when AVF is illuminated by a plane wave. The acoustic pressure field when the UAM is rotated by angle of (c) $\theta_e = 40°$ ( $l^\xi = -1$ ) or (d) $\theta_e = -40°$ ($l^\xi = +1$) while the TC of incident vortex is $l_{in} = +1$.

Fig. 4(a) shows the final phase profile of the AVF under designated rotations, where two opposite phase gradients are accordingly obtained. The diagram and the calculated acoustic pressure field when the AVF being illuminated with plane wave is schematically illustrated in Fig. 4(b), where the rotation of the UAM by angle of ±40°corresponds to the AVF of ITC $ml$. Fig. 4(c) shows the field when setting the rotation angle of the UAM to be 40° to make $l^\xi = -1$. In this case, the incident vortex beam of TC 1 should be converted into a plane wave after passing through the AVF. Then, by changing the rotation angle of the UAM from 40° to –40°, the ITC of AVF becomes $l^\xi = +1$. The full-wave simulation given in Fig. 4(d) shows apparent reflection of the incident vortex by our AVF, and the incident wave is almost prohibited to passing through the device. At the transmission space, the vortex-like phase profile is understood as partially diffracted vortex contributed by higher-order phase gradient of the AVF. Based on the amplitude distribution shown in Fig. 4(c, d), we could observe apparent field intensity contrast when illuminated the AVF with two vortices of opposite

TC, which indicates that the vortex filer functionality could properly work.

Next, we experimentally investigate the vortex filtering functionality of our tunable AVF. Considering the compactness of the AVF device, we only test the AVF sample made of 9 sectors. The PGM and AVF is fabricated with the state-of-art 3D printing technique, whose constituent material is photosensitive resin. Fig. 5(a) and (b) are the cross-sectional image of PGM sample and the photo of AVF sample, respectively. The experimental system is shown in Fig. 5(c) and (d), where the measurement is carried out in a cylindrical waveguide. We employ one speaker as the sound source, after the sound wave transmit through the vortex generation device, i.e., the PGM shown in Fig. 5(a), a vortex beam with a topological charge of 1 can be formed. The AVF is placed in the middle of the two waveguides of inner radius 10 cm and length 110 cm, and the microphone is placed at the tail of the waveguide to collect the signal and we can get the corresponding data from the oscilloscope. The transmission loss (TL) measured by the experiment at different rotation angles is shown in Fig. 5 (e), and it can be observed that when the operating frequency is between 2260 Hz and 2350 Hz, the TL difference between the two rotation angles oscillates around 3 dB, which proofs that the AVF is effective.

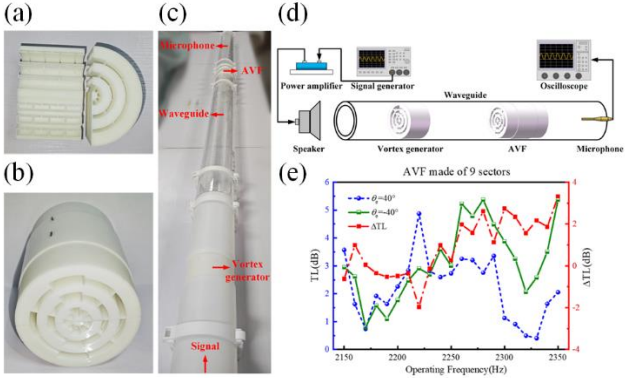

FIG. 5 Experiments for the AVF based on tunable metasurfaces. Photo of the sample (a) PGM, (b) AVF, and (c) experiment setup. (d) Sketch map of the experimental system. (e) Measured TL over the operating frequency ranging from 2150 Hz to 2350 Hz, where the relative orientation angle between UAM and LAM is 40° and −40°.

Asymmetric acoustic wave propagation has attracted significant interest, especially the controllable asymmetric acoustic wave transmission[49] and asymmetric transport of acoustic vortex.[50] Here, the reconfigurable AVF can also realize asymmetric transmission of the acoustic vortex with assigned TC value by simple mechanical rotation. We take the case of $l_M = 2$ and $m = 9$ for illustration, and set $R$ as 8 cm and the operating wavelength $\lambda$ is 15 cm. Considering an acoustic vortex of TC $l_{in} = +2$ that is incident from the left side. When the rotation angle of the UAM is −120°, then the ITC of AVF is $l^\xi = -3$, therefore, the incident vortex is transformed into the acoustic vortex of $l^t = 2 + (-3) = -1$ by the AVF, see Fig. 6 (a). Fig. 6(b) shows the amplitude and phase of the input and transmitted acoustic field on the cut plane at $z = \lambda$. The field distribution at the $z = \lambda$ cross section reveals the generation of $l_{out}^t = -1$ on the transmission. However, when the acoustic vortex of TC $l_{in} = +2$ is incident from the right side, the wave is almost reflected by the AVF, see Fig. 6(c). Fig. 6(d) shows the corresponding amplitude and phase of the acoustic field at the input and transmission side, where the transmission filed intensity is small and close to 0. The transmission efficiency is defined as the ratio of transmitted sound energy to incident sound energy, i.e., $\tau = W_{tran}/W_{inc}$, where $W_{inc}$ and $W_{tran}$ refer to the incident and transmitted sound energy. When an acoustic vortex of TC $l_{in} = +2$ is incident from either the left or right side, the transmission efficiency $\tau$ is 72.0% and 17.2%, respectively.

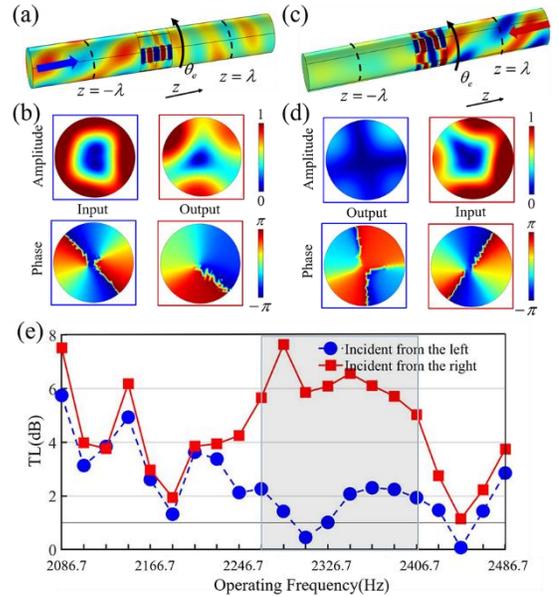

FIG. 6 Asymmetric acoustic vortex transformation via the AVF of which the rotation angle of the UAM is specified as $\theta_e = -120°$. The acoustic field when the vortex of TC $l_{in} = +2$ is incident from the (a) left and (c) right side. The corresponding amplitude and phase of the input and output for the incident vortex wave from the (b) left and (d) right side. (e) TL over the operating frequency ranging from 2086.7 Hz to 2486.7 Hz.

In this work，we utilize the transmission loss (TL), i.e., the decibel difference between the incident and transmitted sound energy, to evaluate the performance of the asymmetric transmission of the AVF. And TL is defined by $TL = 10\log(\frac{W_{inc}}{W_{tran}}) = 10\log(\frac{1}{\tau})$. In our design, the TL of the left-to-right incidence and the right-to-left incidence are 1.43 dB and 7.65 dB, respectively. The TL difference is 6.22 dB, which means the energy transmitted from the right is only 23.8% of the energy transmitted from the left side. The apparent TL difference confirms the asymmetric transmission of the acoustic vortex in AVF. We further investigate the TL over the operating frequency range from 2086.7 Hz to 2486.7 Hz (±200 Hz around the central operating frequency), and the simulation results are shown in Fig. 6(e). Here, we observe obvious asymmetric transmission performance around the central operating wavelength (2246.7 Hz ~ 2406.7 Hz), which indicates that the AVF is still available to filter the right-side incident vortex with the TL contrast greater than 3 dB within about 160 Hz frequency range. Similarly, when we tune the orientation angle of the UAM to be 120°, the ITC of AVF becomes $l^\xi = +3$, and the asymmetric transmission would hold true for the vortex of TC $l_{in} = -2$. Therefore, our AVF could support tunable asymmetric transmission of the OAM via simple mechanical rotation.

In conclusion, we have proposed an acoustic dual-layer metasurface for tunable and selective filtering of the incident acoustic vortex, which provides an external DoF to extract the information carried by the desired OAM channels. Based on the complex vortex diffraction supported by the acoustic gradient metasurface, selective transmission of OAM can be achieved by simply mechanically rotating the UAM at specific angles. This rotation alters the ITC of the AVF, which facilitates the control of the critical order of vortex and thereby reflects the incident vortex. We theoretically demonstrate the selective filtering functionality for the vortex beams of opposite TCs. In addition, such OAM selective transmission property could be further applied to asymmetric acoustic beam transmission, which endows the possibility for the customized OAM one-way transport. In the future, our metasurface-based OAM filter strategy has the potential for flexible channel switching in the OAM-based underwater acoustic communication.

This work was supported by the National Natural Science Foundation of China (Grant No. 61775050, 12104044).

## SUPPLEMENTARY MATERIAL

In Supplementary Material, we add detailed discussion on the phase profile of the UAM and LAM, diffraction property of ideal AVF, analysis on the performance of AVF made of realistic structures, performance of AVF when illustrated by vortex generated by realistic PGM, and influence of thermal acoustic effect on the vortex filtering performance of our AVF.

## AUTHOR DECLARATIONS

### Conflict of Interest

The authors have no conflicts to disclose.

## DATA AVAILABILITY

The data that support the findings of this study are available from the corresponding author upon reasonable request.